\documentclass[11pt,amsmath,amssymb,amsfonts,superscriptaddress,floatfix,showpacs]{revtex4-1}

\usepackage{latexsym}
\usepackage{graphics}
\usepackage{graphicx}
\usepackage[dvips]{color}
\usepackage{amsmath}
\usepackage[english]{babel}

\begin{document}
\newcommand{\red}{\color{red}}

\title{Hybrid density functional study of electronic and optical properties 
of phase change memory material: $\mathrm{Ge_{2}Sb_{2}Te_{5}}$}

\author{T. Kaewmaraya}
\affiliation{Department of Physics and Astronomy, Box 530, Uppsala
  University, S-751 21 Uppsala, Sweden}
\author{M. Ramzan}
\affiliation{Department of Physics and Astronomy, Box 530, Uppsala
  University, S-751 21 Uppsala, Sweden}
\author{H. L\"{o}f\aa s}
\affiliation{Department of Physics and Astronomy, Box 530, Uppsala
  University, S-751 21 Uppsala, Sweden}
\author{Rajeev Ahuja} 
\affiliation{Department of Physics and Astronomy, Box 530, Uppsala
  University, S-751 21 Uppsala, Sweden} \affiliation{Department of
  Materials and Engineering, Royal Institute of Technology (KTH),
  S-100 44 Stockholm, Sweden}

\begin{abstract} { In this article, we use hybrid density functional (HSE06) 
to study the crystal and electronic structures and optical properties of 
well known phase change memory material $\mathrm{Ge_{2}Sb_{2}Te_{5}}$. 
We calculate the structural parameters, band gaps and dielectric functions 
of three stable structures of this material. We also analyze the electron 
charge distribution using the Bader's theory of charge analysis. We find 
that hybrid density functional slightly overestimate the value of 'C' 
parameter. However, overall, our results calculated with the use of 
hybrid density functional (HSE06) are very close to available 
experimental values than calculated with the use of PBE functional. 
Specifically, the electronic band gap values of this material calculated 
with HSE06 are in good agreement with the available experimental 
data in the literature. Furthermore, we perform the charge analysis 
and find that naive ionic model fails to explain the charge distribution 
between the constituent atoms, showing the complex nature of this compound.}
\end{abstract}

\maketitle

\section{Introduction}
$\mathrm{Ge_{2}Sb_{2}Te_{5}}$ (GST) chalogenides have been of technological
importance due to their applications for rewritable data-storage devices 
such as compact-disk (CD) and digital versatile disk (DVD). 
The underlying principle behind the data-storage process is based on the 
fast and reversible phase transformation between a crystalline phase and 
an amorphous phase, leading to changes in electrical conductivity 
and optical reflectivity~\cite{Ovshinsky,Libera,Yamada_MRS}. In particular, 
the phase change happens at relatively low temperature range
\cite{Yamada_1991,Friedrich_2000,Lee}, 
making it feasible for phase change random access memory (PCRAM). 
More physical understanding of both amorphous and crystalline 
GST have been essentially needed since it could guide us to further
enhancement.  

The GST crystalline phase consists of two states namely metastable
phase and stable phase. Based on high-resolution electron microscopy analysis, 
the metastable phase has been proposed to crystallize in the rocksalt-type 
structure(Fm$\overline{3}$m) that Te atoms completely occupy the 4(a) site 
and Ge, Sb, and intrinsic vacancies randomly atoms completely occupy 
the 4(b) site\cite{Park_JAP_2005}. 
On the other hand, the stable phase crystallizes in the hexagonal structure. 
Three different atomic arrangements have been proposed. 
I. I. Petrov \textit{et al.} \cite{Petrov} firstly performed an 
experiment by using electron 
transmission microscope and they reported that GST is the hexagonal 
structure with space group P$\overline{3}$m1 having corresponding
lattice constants a = 4.20 \AA \,\, and c = 16.96 \AA, respectively. 
Te atoms occupy 1(a), 2(d), and 2(d) sites while Sb and Ge atoms occupy 
2(d) and 2(c) sites. Later, B. J. J. Kooi \textit{et al.}\cite{Kooi} argued 
that Ge atoms occupy 2(d) sites and Sb atoms occupy 2(c) sites. 
However, T. Matsunaga \textit{et al.}\cite{Matsunaga} further investigated 
this by x-ray diffraction and they found different results. 
According to them, GST crystallizes 
in hexagonal structure with the space group P$\overline{3}$m1 and the lattice 
parameters a = 4.2247 \AA \,\,and c = 17.2391 \AA. They have 
also indicated that Ge and Sb atoms randomly occupy 2(d) and 2(c) sites. 
Up to now, complete explanations of atomic arrangement of the GST stable 
phase has remained unclear. However, B. S. Lee and co-workers \cite{Lee}
 and J. W. Park \textit{et al.} \cite{Park}
experimentally 
studied the electronic and optical properties of the stable GST phase. 
In addition, there are also 
theoretical investigations related to the stable GST phase. 
Z. Sun \textit{et al.}\cite{Sun_2006} carried out first-principles electronic
structure 
calculations based on the density functional theory(DFT) to compare three
proposed models and they have concluded that the configuration proposed 
by Kooi \textit{et al.}\cite{Kooi} is the most stable one. 

Hybrid functional of J. H. Heyd, G. E. Scuseria and M. Ernzerhof, 
better called HSE06 \cite{Heyd} has been shown to give improved structural 
parameters for a number of systems as compared to the local density 
approximations (LDA) and Generalized gradient approximation (GGA)
\cite{Paier,Marsman}. In addition, they also provide improved band gaps 
which are close to experiment, slightly lower for most cases
\cite{Paier,Marsman}. Up to now, no theoretical studies have been 
conducted to investigate the structural and optical properties of 
GST material by using 
hybrid density functionals. In this work, we therefore perform the calculations 
to address structural and electronic properties of stable GST using GGA 
and hybrid density functional (HSE06). The optimized structural parameters and 
electronic structures are presented.  Finally, the optical properties
of this compound are also investigated. 

\section{Methods/Computational details}

\textit{Ab-initio} total energy calculations based on the density functional 
theory(DFT)\cite{Kohn} and all-electron projector-augmented wave 
method\cite{Blochl} have been performed by using the VIENNA AB INITIO 
SIMULATION PACKAGE (VASP)\cite{Kresse94,Kresse99}. The atomic 
structures were constructed according to the experimental data provided by 
refs. \cite{Petrov,Kooi,Matsunaga}. The generalized gradient 
approximation of Perdew Burke-Ernzerhof (PBE)\cite{PBE} was employed 
as exchange-correlation 
functionals. 14 electrons ($3d^{10}4s^{2}4p^{2}$) of Ge, 5 electrons 
($5s^{2}5p^{3}$) of Sb, and 6 electron ($5s^{2}5p^{4}$) of Te were 
treated as valence electrons in the pseudopotentials. 
The cutoff energy for plane wave basis of 800 
eV and the k-point mesh for brillouin zone integration of 8x8x2 were 
used since they provide sufficient convergence in total energy of structural
optimization. On the other hand, the denser k-point mesh of 16x16x8
was adopted for calculating density of states(DOS) and dielectric
functions.

Calculations using hybrid density functional (HSE06)\cite{Heyd} 
were also comparatively carried out. In this particular case, 
the exchange-correlation functionals are the rational mixing between the 
Fock exchange, PBE exchange and PBE correlation
\begin{equation}  
\textrm{$E^{HSE}_{xc}$ = 1/4 $E^{HF,SR}_{x}$($\mu$) + 
3/4 $E^{PBE,SR}_{x}$($\mu$) + $E^{PBE,LR}_{x}$($\mu$) + $E^{PBE}_{c}$},
\end{equation}
The PBE exchange term is decomposed into two parts namely short range (sr) 
and long-range (lr) while the correlation part is totally from PBE. 
The parameter $\mu$ represents the range when the short range term is 
negligible and it is $0.207^{-1}$ for HSE06. The detailed mathematical 
derivations and tests of HSE06 functionals are given in ref. 
\cite{Heyd}. In order to obtain satisfactory results with reasonable 
computing time, the lower cutoff energy and k-point mesh, 600 eV and 4x4x2, 
respectively were used for structural optimization. The denser k-point mesh
was adjusted higher to be 8x8x2 for calculating DOS and dielectric functions.
The conjugate gradient scheme utilized for electronic relaxation algorithm 
is applied to all the structural optimization. 
The volume, shape, and atomic positions were fully optimized and relaxations 
were allowed until the Hellmann-Feynman forces on the atoms were less than 
$0.005$ eV/\AA. In addition, the calculated electronic charge density 
from the optimized atomic structure was used to calculate electronic 
charge partitioned for each atom by using a grid-based Bader charge 
analysis \cite{Tang,Sanville}. The zero flux  surface of the 
electronic charge density is used to determine the amount of charge occupied
by that particular atom.

\section{Results and discussion}

\begin{figure}[h!]
\includegraphics[width=10cm]{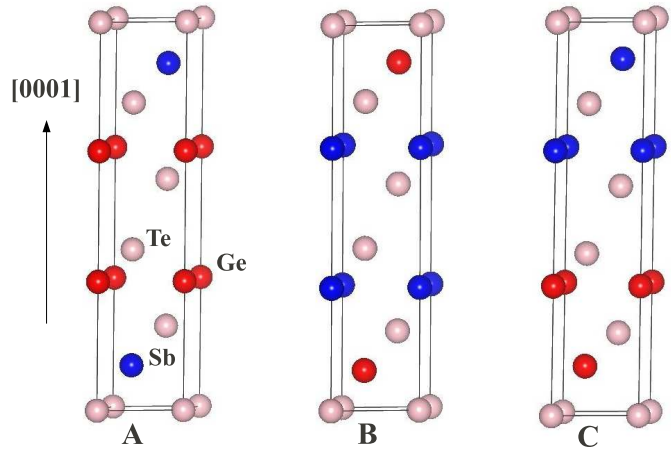}
\caption{
The optimized atomic structures of $\mathrm{Ge_{2}Sb_{2}Te_{5}}$
    proposed by (A) I. I. Petrov \textit{et al.}\cite{Petrov}, (B) B. J. Kooi \textit{et. al.}\cite{Kooi} and 
(C) T. Matsunaga \textit{et al.}\cite{Matsunaga}.  
    Red, blue, and pink spheres repersent Ge, Sb and Te atoms, respectively. 
 \label{structure}
}
\end{figure}

We start our calculations by optimizing atomic structures of the stable phase 
of GST proposed in references \cite{Petrov,Kooi,Matsunaga} and they 
are labeled as A (I. I. Petrov \textit{et al.}\cite{Petrov}), B (B. J. Kooi 
\textit{et. al.}\cite{Kooi}), and C (T. Matsunaga \textit{et al.}
\cite{Matsunaga}), respectively. The equilibrium 
geometries by using PBE and HSE06 functionals are given in 
\ref{structure}. It can be seen that the unit cell (1 formula unit) consists 
of 9 atoms (2 Ge, 2 Sb and 5 Te). The stable GST is a layered 
structure that Ge, Sb, and Te atoms are stacked along $c$-axis (in the [0001]
direction). The optimized lattice parameters are listed in \ref{tab1}. 
PBE functionals obviously overestimate lattice parameters 
 with the maximum 2 \% difference as 
compared to the reported experimental values. 
This comes from the well-known deficiency that GGA 
functionals overestimate lattice constants. 
However, we have found that
our calculations are in good agreement with those previously reported
calculations as also indicated in \ref{tab1}. 
However, using the HSE06 functional, the lattice constant $a$ is improved 
to be closer to the experimental values, but the lattice constant
$c$ is overestimated. This could be related to
that the stable GST is a layered structure along $c$-axis. 
The interactions between adjacent layers are probably low and 
hybrid functional may fail to explain these weak interactions.

\begin{figure}[h!]
\includegraphics[width=10cm]{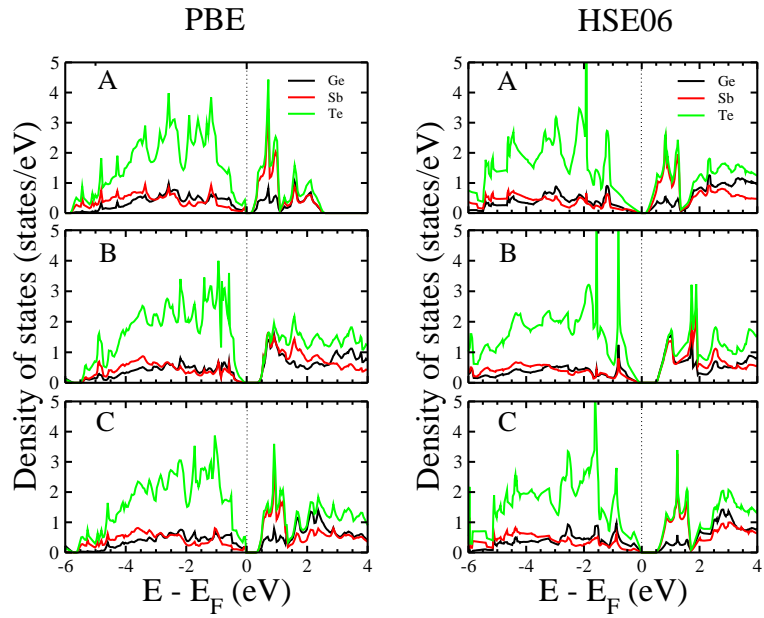}
\caption{Density of states (DOS) of $\mathrm{Ge_{2}Sb_{2}Te_{5}}$ calculated with
   PBE (left side) and HSE06 (right side). The fermi level
   is shifted to zero.}
\label{DOS}
\end{figure}

   After acquiring the complete information about structural 
parameters, we proceed to investigate the corresponding electronic 
structures of these stable phases of $\mathrm{Ge_{2}Sb_{2}Te_{5}}$. 
Our calculated density of states (DOS) with the use of PBE and HSE06 
functionals of these stable phases are shown in \ref{DOS}. We see 
that PBE and HSE06 give almost similar DOS and it is found 
that Te-derived states are mainly formed at the top of the valence band 
while  Ge, Sb, and Te share the bottom of conduction band. However, 
there is strong hybridization of Sb and Te 
atoms in the conduction band around 0.5-1.0 eV for A and C whereas Ge, Sb,
and Te atoms almost equally hybridize for B. This can be explained by their
different atomic arrangements. For A, Sb atoms from 2(d) sites are 
surrounded by Te atoms from 1(a), 2(d), and 2(d) sites.
For B, Sb atoms from 2(c) site are surrounded by Te atoms from 1(a) site.
In addition,  states in the conduction band have very similar shapes but 
they are pushed upward, resulting in higher band gaps as indicated
in \ref{table:2}. PBE functional predicts band gaps as 0.00 eV, 0.24 eV, 
and 0.22 eV for A, B, and C, respectively while it has been reported that 
stable GST has the band gap ranging from 0.50-0.57 eV\cite{Park,Lee}.
The band gaps are quantitatively underestimated because of the main 
deficiency of DFT to deal with excitation. However, HSE06 hybrid 
functional gives band gaps closer to the experimental values found 
in the literature. 
The C phase has the band band gap of 0.48 eV which is in good 
agreement with the experimental
values while B phase has slightly lower, 0.37 eV.
On the other hand, the phase A has the smallest band gap
of 0.26 eV.

\begin{figure}[h!]
 \begin{center}
  \includegraphics[width=10cm]{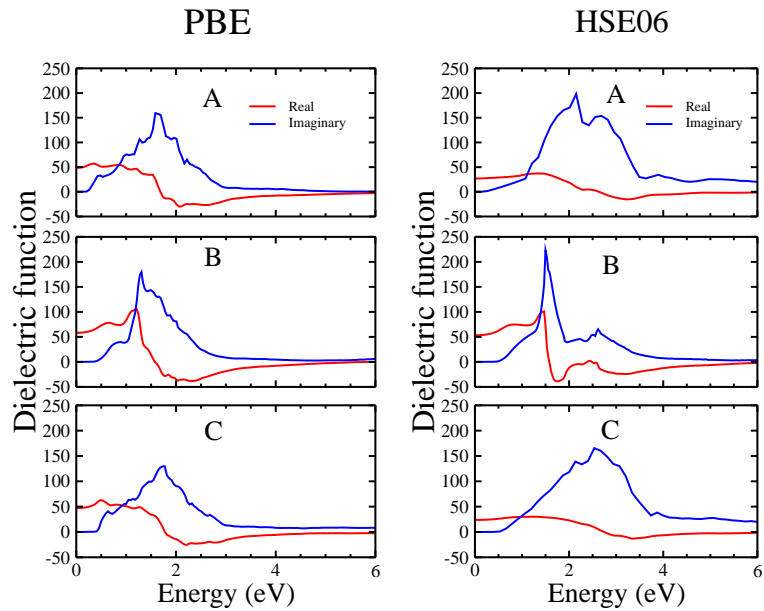}
 \caption{The real and imaginary parts of dielectric 
           functions of $\mathrm{Ge_{2}Sb_{2}Te_{5}}$
           calculated with PBE (left side) and HSE06 (right side).}
\label{Dielectric} 
   \end{center}
\end{figure}

In \ref{Dielectric}, we show the real and imaginary parts of dielectric
functions of $\mathrm{Ge_{2}Sb_{2}Te_{5}}$. The left side represents 
the dielectric functions calculated by
using PBE functional whereas the right side denotes those calculated by using
hybrid density functional. For PBE functional, A and C have nearly similar
dielectric functions. They have the same amplitude of the real part dielectric 
function at zero energy and their main peaks of the imaginary part are located
between 1.5-2.0 eV, in good agreement with reported results in 
ref. \cite{Park}. However, the main peak of the imaginary part of B is slightly
lower than A and C. It has been experimentally reported that the imaginary 
dielectric function locates at 1.5 eV\cite{Park}. For 
HSE06 functional, It is found that A, B, and C, have slightly different 
dielectric functions considering in terms of the locations of main peaks 
of the imaginary part, the starting point at zero energy of the real part 
and the amplitudes.

We have also calculated the electronic charge distribution by using Bader
Charge analysis\cite{Tang,Sanville} and list our results in \ref{table:3}. 
As mentioned earlier that Ge, Sb, and Te have 14, 5, and 6 valence 
electrons, respectively. In a pure ionic model, Ge and Sb would loose 
4 and 3 electrons respectively and Te may gain $\approx$ 3 electrons.
But in our case the situation is rather complex. For instance in phase A, 
Ge and Sb are loosing only 0.31 and 0.6 electrons respectively,
which are gained by Te. These results represent the complex nature of 
this material and reveal the importance of quantitative analysis
 over a simple ionic model. In the three proposed stable phases of 
this material, the charge distribution is almost same. 
Although, the values obtained for B phase favors slightly the charge 
transfer from Ge and Sb to Te, but the numbers are not fundamentally 
different from those, which were calculated for A and C structures.    

 \section{Conclusions}

In summary, we have performed comparative study of the structural, 
electronic and optical properties of the stable structures of 
$\mathrm{Ge_{2}Sb_{2}Te_{5}}$, with the use of GGA and hybrid density 
(HSE06) functionals. In our present study, we have shown that structural 
parameters and electronic band gap of this material 
calculated with the use of hybrid density functional (HSE06) are in 
good agreement with the available experimental results than calculated 
with the use of PBE functional. However, HSE06 functional slightly 
overestimate the C parameter and optical properties of this compound. 
We have also analyzed the charge distribution between the 
constituent elements of this material using the Bader's theory 
of atoms and we find that, due to the complex nature of this compound, 
the simple ionic model fails to explain it. We have shown that , on the overall, 
hybrid density functional (HSE06) is important for the correct description 
of GST material and especially to reproduce the electronic structure
of this compound. Finally, we have presented that all the calculated 
parameters of stable phase B of $\mathrm{Ge_{2}Sb_{2}Te_{5}}$ are 
more closer to available experimental data than stable phases A and C.


We would like to acknowledge VR and FORMAS for
financial support. T.K.\ would also like to acknowledge the Royal Thai 
Government for financial support. M. Ramzan acknowledges 
Higher Education Commission of Pakistan. SNIC and SNAC 
have provided computing time for this project. 

%


\newpage
\begin{table}[h!]
\caption{The calculated structural parameters of the stable 
         $\mathrm{Ge_{2}Sb_{2}Te_{5}}$ phase.}
\begin{center}
\bigskip                          
 \begin{tabular}{llccc } \hline\hline
 Proposed structures & $xc$-functionals & $a_{0}$ (\AA) & $c_{0}$ (\AA)    \\ \hline
  A             & PBE               & 4.263  & 17.971 &   \\
                & PBE$^{e}$         & 4.270  & 17.720 &   \\
                & HSE06             & 4.216  & 17.832 &   \\
                & Experiment$^{a}$  & 4.200  & 16.960 &   \\
  B             & PBE               & 4.300  & 17.237 &   \\
                & PBE$^{e}$         & 4.290  & 17.250 &   \\
                & HSE06             & 4.230  & 18.058 &   \\
                & Experiment$^{b}$  & 4.200  & 16.960 &   \\
  C             & PBE               & 4.278  & 17.743 &   \\
                & PBE$^{d}$         & 4.270  & 17.890 &   \\
                & HSE06             & 4.204  & 18.196 &   \\
                & Experiment$^{c}$  & 4.220  & 17.240 &   \\ \hline \hline
                $^{a}$Experiment in Ref. \cite{Petrov}.  &    &     & \\
                $^{b}$Experiment in Ref. \cite{Kooi}.    &    &     & \\
                $^{c}$Experiment in Ref. \cite{Matsunaga}. &    &   &  \\
                $^{d}$DFT culculations in Ref. \cite{Park}. &    &   & \\
                $^{e}$DFT culculations in Ref. \cite{Kim}. &    &   & \\
              
\end{tabular}
\end{center}
\label{tab1}
\end{table}

\begin{table}[h!]
\caption{The calculated band gap energies of the 
         stable $\mathrm{Ge_{2}Sb_{2}Te_{5}}$ phase.}
\begin{center}
\bigskip
\begin{tabular}{ccc } \hline\hline
 Proposed structures & PBE (eV) & HSE06 (eV)    \\ \hline
  A         &  0.00    &  0.26  \\
  B         &  0.24    &  0.37  \\
  C         &  0.22    &  0.48  \\ \hline 
 Experiments     & \multicolumn{2}{c}{0.57 (Ref. \cite{Park}),
 0.50 (Refs. \cite{Lee})} \\ \hline \hline
\end{tabular}
\end{center}
\label{table:2}
\end{table}

\begin{table}[h!]
\caption{The calculated electronic charges of the stable 
          $\mathrm{Ge_{2}Sb_{2}Te_{5}}$ structures by Bader charge analysis.}
\begin{center}
\bigskip
\begin{tabular}{cccc } \hline\hline
                & \multicolumn{3}{c}{electronic charge ($e$)}  \\
                \cline{2-4}
 atomic species & A   & B    & C     \\ \hline
 Ge(1)          & 13.69  & 13.60 &  13.66 \\
 Ge(2)          & 13.69  & 13.60 &  13.66 \\
 Sb(1)          &  4.40  &  4.35 &   4.40 \\
 Sb(2)          &  4.40  &  4.35 &   4.40 \\
 Te(1)          &  6.52  &  6.44 &   6.49 \\
 Te(2)          &  6.29  &  6.37 &   6.32 \\
 Te(3)          &  6.29  &  6.37 &   6.32 \\
 Te(4)          &  6.36  &  6.46 &   6.37 \\
 Te(5)          &  6.36  &  6.46 &   6.37 \\ \hline \hline
\end{tabular}
\end{center}
\label{table:3}
\end{table}

\newpage

\newpage

\newpage


\end{document}